\documentclass[english]{elsart}
\usepackage{times}
\usepackage[T1]{fontenc}
\usepackage[latin1]{inputenc}
\usepackage{graphicx}

\makeatletter

\providecommand{\LyX}{L\kern-.1667em\lower.25em\hbox{Y}\kern-.125emX\@}

\usepackage{babel}
\makeatother
\begin{document}
\begin{frontmatter}

\title{Carbon Doping in $MgB_{2}$ : Role of Boron and Carbon $p_{x(y)}$
Bands}

\author{Prabhakar P. Singh}


\address{Department of Physics. Indian Institute of Technology, Powai, Mumbai-400076,
India }

\begin{abstract}
We have studied the changes in the electronic structure and the superconducting
transition temperature $T_{c}$ of $Mg(B_{1-x}C_{x})_{2}$ alloys
as a function of $x$ with $0\leq x\leq 0.3$. Our density-functional-based
approach uses coherent-potential approximation to describe the effects
of disorder, Gaspari-Gyorffy formalism to estimate the electron-phonon
matrix elements and Allen-Dynes equation to calculate $T_{c}$ in
these alloys. We find that the changes in the electronic structure
of $Mg(B_{1-x}C_{x})_{2}$ alloys , especially near the Fermi energy
$E_{F}$, come mainly from the outward movement of $E_{F}$ with increasing
$x$,  and the effects of disorder in the $B$ plane are small. In
particular, our results show a sharp decline in both $B$ and $C$
$p_{x(y)}$ states for $0.2\leq x\leq 0.3$. Our calculated variation
in $T_{c}$ of $Mg(B_{1-x}C_{x})_{2}$ alloys is in qualitative agreement
with the experiments.
\end{abstract}

\end{frontmatter}

\section{INTRODUCTION}

The changes in the electronic properties of $MgB_{2}$ upon doping
with various elements gets manifested in the changes in its superconducting
behavior. An understanding of these changes in the superconducting
behavior of $MgB_{2}$ upon doping can provide useful information
about the nature of interaction responsible for superconductivity.
Such information can then be used to prepare materials with desired
superconducting properties. Since the discovery of superconductivity
in $MgB_{2}$, a substantial amount of effort has been directed towards
probing the various aspects of superconductivity using doping. 

The effects of doping $MgB_{2}$ with various elements such as $s$-elements
$Be$ \cite{felner} and $Li$ \cite{zhao}, $sp$-elements $Al$
\cite{xiang,slusky} and $Si$ \cite{cimberle}, and $d$-elements
$Fe,\, Co,\, Ni$ \cite{moritomo} and others \cite{buzea} have been
studied experimentally. Most of these substitutions take place in
the $Mg$ plane of $MgB_{2}$, leaving the $B$ plane in tact. For
some dopants at particular concentrations, it is also found that the
dopants prefer to form a separate layer \cite{xiang,slusky} rather
than mix with the $Mg$ layer or the $B$ layer. For dilute amount
of dopants the changes in the superconducting properties, in particular
the superconducting transition temperature, $T_{c}$, depends on the
alloying element \cite{felner,zhao,xiang,slusky,cimberle,moritomo,buzea}.
For larger concentrations, with only $Mg_{1-x}Al_{x}B_{2}$ \cite{xiang,slusky}
being studied extensively, the $T_{c}$ is found to decrease due to
the gradual filling of the $B$ $p_{x(y)}$-holes.

The doping of $MgB_{2}$ with $C$ \cite{ahn,paran,takenobu,zhang,bharathi,ribeiro}
can be expected to be different from the ones described in the previous
paragraph due to the following two reasons, (i) C prefers to go to
the $B$ plane instead of the $Mg$ plane and (ii) C forms strong
covalent-bonded layered structure similar to $B$ layer in $MgB_{2}$.
These reasons imply that the lowering of $T_{c}$ in $Mg(B_{1-x}C_{x})_{2}$
alloys with increasing $x$ is mainly due to the filling up of the
$p_{x(y)}$ holes, and not due to disorder in the $B$ plane. However,
the $x$ dependence of $T_{c}$ in $Mg(B_{1-x}C_{x})_{2}$ alloys
as obtained in some earlier experiments may not be accurate due to
the difficulties in estimating its $C$ content \cite{ribeiro}. According
to Ref. \cite{ribeiro}, the $T_{c}$ observed for $x=0.1$ in $Mg(B_{1-x}C_{x})_{2}$
alloys was $22\, K$, which is lower than the result of Ref. \cite{bharathi}. 

Theoretical attempts at understanding the changes in the electronic
structure and how these changes affect the superconducting properties
of carbon-doped $MgB_{2}$ alloys have not gone beyond the rigid-band
model approach \cite{mehl} with its limited applicability. To be
able to reliably describe the effects of $C$ doping in $MgB_{2}$,
we have carried out \textit{ab initio} studies of $Mg(B_{1-x}C_{x})_{2}$
alloys with $0\leq x\leq 0.3$. We have used Korringa-Kohn-Rostoker
coherent-potential approximation \cite{pps_cpa,faulkner} in the atomic-sphere
approximation (KKR-ASA CPA) method for taking into account the effects
of disorder, Gaspari-Gyorffy formalism \cite{gaspari} for calculating
the electron-phonon coupling constant $\lambda $, and Allen-Dynes
equation \cite{allen1} for calculating $T_{c}$ in $Mg(B_{1-x}C_{x})_{2}$
alloys as a function of $C$ concentration. We have used the present
approach to study several other $MgB_{2}$-based alloys \cite{pps_mgtab2,pps_mgalb2,pps_3dmgb2}.
In the following, we present our results in terms of the changes in
the total density of states (DOS), in particular the changes in the
$p$ contributions of both $B$ and $C$ to the total DOS, as a function
of concentration $x$ of $C$ atoms. Before we describe the results
of our calculations we provide some of the computational details.

\section{COMPUTATIONAL DETAILS}

The charge self-consistent electronic structure of $Mg(B_{1-x}C_{x})_{2}$
alloys as a function of $x$ has been calculated using the KKR-ASA
CPA method \cite{pps_cpa,faulkner}. We parametrized the exchange-correlation
potential as suggested by Perdew-Wang \cite{perdew1} within the generalized
gradient approximation \cite{perdew2}. The Brillouin zone (BZ) integration
was carried out using $1215$ $k-$ points in the irreducible part
of the BZ. For DOS calculations, we added a small imaginary component
of $1$ $mRy$ to the energy and used $4900$ \textbf{k}-points in
the irreducible part of the BZ. The lattice constants for $Mg(B_{1-x}C_{x})_{2}$
alloys as a function of $x$ were taken from the experimental result
of . The Wigner- Seitz radius for $Mg$ was larger than that of $B$
and $C$. The sphere overlap which is crucial in ASA, was less than
$10$\% and the maximum $l$ used was $l_{max}$ = $2$. The present
calculations were carried out using the scalar-relativistic Schroedinger
equation. The Green's function is calculated in a complex plane with
a $20$ point Gaussian quadrature.

The electron-phonon coupling constant $\lambda $ was calculated using
Gaspari-Gyorffy \cite{gaspari} formalism with the charge self-consistent
potentials of $Mg(B_{1-x}C_{x})_{2}$ obtained with the KKR-ASA CPA
method. Subsequently, the variation of $T_{c}$ as a function of $C$
concentration was calculated using Allen-Dynes equation \cite{allen1}.
The parameters used in the calculation of $T_{c}$ using the Allen-Dynes
equation, for the whole range of $x$, were $\mu ^{*}=0.09$, $\omega _{rms}=400\, cm^{-1}$,
and $\omega _{ln}=834\, K$.

\section{RESULTS AND DISCUSSION}

In this section we describe our results for $Mg(B_{1-x}C_{x})_{2}$
alloys in terms of the changes in the densities of states, the electron-phonon
coupling constant $\lambda $ and the superconducting transition temperature
$T_{c}$ as a function of $x$ with $0\leq x\leq 0.3$. In our analysis
we will emphasize those aspects of changes in the electronic structure
which directly affect the superconducting properties of $Mg(B_{1-x}C_{x})_{2}$
alloys, in particular its $T_{c}$. Thus, we discuss the changes in
the density of states as a function of $x$, in particular the $p$
contributions of both $B$ and C to the total DOS and how it affects
the electron-phonon coupling constant and consequently the $T_{c}$
of $Mg(B_{1-x}C_{x})_{2}$ alloys.

\subsection{Densities of States}

In Fig. 1, we show the total DOS of $Mg(B_{1-x}C_{x})_{2}$ alloys
for $x$ equal to $0.01,$ $0.1,$ $0.2$ and $0.3$, calculated using
the KKR-ASA CPA method. A comparison of Figs. 1(a)-(d) shows that
with increasing $x$ the Fermi energy, $E_{F},$ moves outward and
the DOS becomes smoother. The outward movement of $E_{F}$ is due
to the increase in the number of valence electrons with the addition
of $C$ while the increase in disorder in the $B$ plane of $MgB_{2}$
flattens the peaks in the DOS. In $Mg(B_{1-x}C_{x})_{2}$ alloys the
outward movement of $E_{F}$ decreases the total DOS at the Fermi
energy, $N(E_{F})$, in particular for $x\sim 0.3$, $N(E_{F})$ is
a minimum. Generally, a decrease in $N(E_{F})$ leads to a lowering
of $T_{c}$ , it is expected that with the increase in $C$ concentration
the $T_{c}$ of $Mg(B_{1-x}C_{x})_{2}$ alloys should decrease. 

\begin{figure}
\begin{center}\includegraphics[  width=7.4cm,
  angle=270,
  origin=lB]{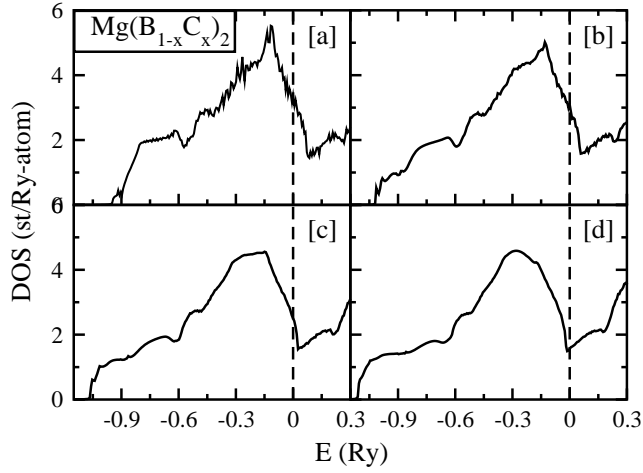}\end{center}

\caption{The calculated total density of states (solid line) of $Mg(B_{1-x}C_{x})_{2}$
alloys for (a) $x=0.01$, (b) $x=0.1$, (c) $x=0.2$ and (d) $x=0.3$,
respectively. The dashed vertical line indicates the Fermi energy. }
\end{figure}

It is known that in $MgB_{2}$ the holes in the $p_{x(y)}$ bands
couple very strongly to the optical $E_{2g}$ phonon mode \cite{kong,bohnen,choi,pps_nbb2},
and hence a change in the $p_{x(y)}$ bands directly affects the superconducting
properties of $MgB_{2}$. Thus, to get a more detailed picture of
how the changes in $N(E_{F})$ affect the superconducting properties
of $MgB_{2}$, we have to analyze the changes in the $B$ $p_{x(y)}$
contributions to $N(E_{F})$. In addition, since $C$ is known to
form strong covalent-bonded layered structures, it may be that in
$Mg(B_{1-x}C_{x})_{2}$ alloys the $C$ $p_{x(y)}$ electrons/holes
also couple to the optical phonons the way $B$ $p_{x(y)}$ electrons/holes
do. 

\begin{figure}
\begin{center}\includegraphics[  width=7.4cm,
  angle=270,
  origin=lB]{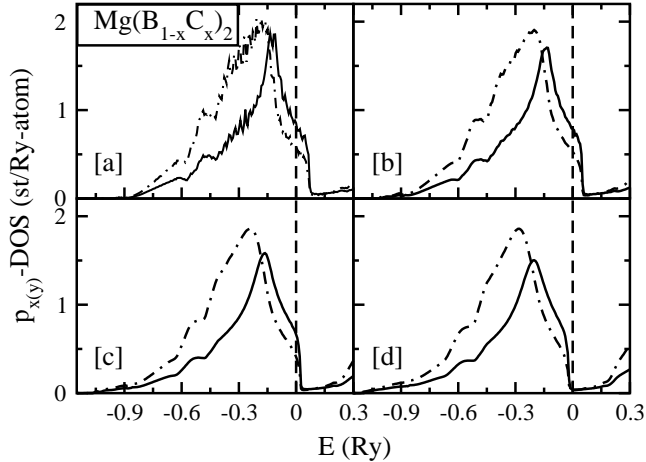}\end{center}

\caption{The calculated $p_{x(y)}$ density of states for both $B$ (solid
line) and $C$ (dot-dashed line) in $Mg(B_{1-x}C_{x})_{2}$ alloys
at (a) $x=0.01$, (b) $x=0.1$, (c) $x=0.2$ and (d) $x=0.3$, respectively.
The dashed vertical line indicates the Fermi energy. }
\end{figure}

To see how the $p$ contributions to the total DOS change in $Mg(B_{1-x}C_{x})_{2}$
alloys as a function of $x$, we show in Fig. 2 the $p_{x(y)}$ DOS
for both $B$ and $C$ in $Mg(B_{1-x}C_{x})_{2}$ alloys for $x$
equal to $0.01,$ $0.1,$ $0.2$ and $0.3$. The $p$ DOS of $MgB_{2}$
calculated with the KKR-ASA CPA method is in good agreement with our
full-potential result \cite{pps_mgb2_prl} and other calculations
\cite{kortus,zhu}, indicating the reliability of the present approach.
From Fig. 2 it is clear that near $E_{F}$ both $B$ and $C$ $p_{x(y)}$
densities of states are similar for the whole range of $x$ with $C$
DOS being somewhat smaller than the $B$ DOS. As the addition of $C$
increases the electron count, the outward movement of $E_{F}$ as
$x$ increases fills up the $p_{x(y)}$bands of both $B$ and C, signalling
a sharp change in the superconducting properties of $Mg(B_{1-x}C_{x})_{2}$
alloys in this range.

\begin{figure}
\begin{center}\includegraphics[  width=7.4cm,
  angle=270,
  origin=lB]{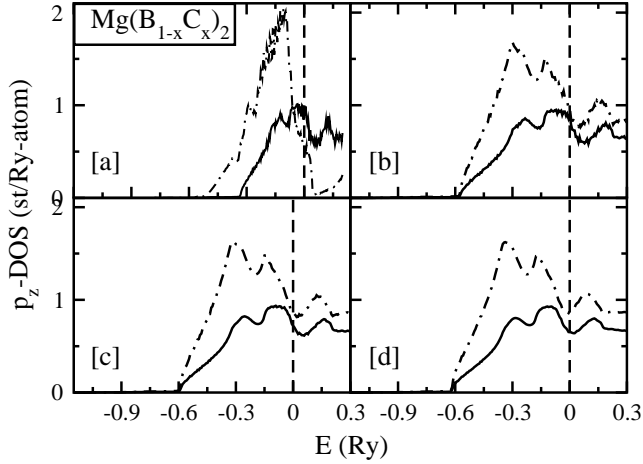}\end{center}

\caption{The calculated $p_{z}$ density of states for both $B$ (solid line)
and $C$ (dot-dashed line) in $Mg(B_{1-x}C_{x})_{2}$ alloys at (a)
$x=0.01$, (b) $x=0.1$, (c) $x=0.2$ and (d) $x=0.3$, respectively.
The dashed vertical line indicates the Fermi energy. }
\end{figure}

The changes in the $p_{z}$ contribution to the total DOS in $Mg(B_{1-x}C_{x})_{2}$
alloys as a function of $x$ is not very significant, as can be seen
from Fig. 3 , where we have shown the $p_{z}$ DOS for both $B$ and
$C$ in $Mg(B_{1-x}C_{x})_{2}$ alloys for $x$ equal to $0.01,$
$0.1,$ $0.2$ and $0.3$. In the dilute limit, the $p_{z}$ DOS due
to $B$ in $Mg(B_{1-x}C_{x})_{2}$ alloy is similar to the $p_{z}$
DOS due to $B$ in $MgB_{2}$ \cite{pps_mgb2_prl,zhu,klie}. In addition,
these electrons play a minor role in deciding the superconducting
properties of $Mg(B_{1-x}C_{x})_{2}$ alloys.

Since the changes in the superconducting properties of $Mg(B_{1-x}C_{x})_{2}$
are dictated by the DOS at $E_{F}$, in particular the $p_{x(y)}$
contributions of both $B$ and $\textrm{C }$ to the total DOS, we
show in Fig. 4 the variations of the total density of states $N(E_{F})$
and the $B$ and $\textrm{C }$ $p_{x(y)}$ contributions to it. As
can be seen from Fig. 4, there is a steep drop in the $p_{x(y)}$
DOS for both $B$ and $C$ between $x=0.2$ to $x=0.3$, indicating
the filling up of the $p_{x(y)}$ bands. Given that the $p_{x(y)}$
bands are responsible for driving superconductivity in $MgB_{2}$,
we expect the superconducting properties of $Mg(B_{1-x}C_{x})_{2}$
alloys to change significantly between $x=0.2$ to $x=0.3$. Here,
we like to point out that the range of $x$ as indicated above, in
which the superconducting properties of $Mg(B_{1-x}C_{x})_{2}$ are
expected to change significantly, must be viewed within the constraints
inherent in the present approach. 

\begin{figure}
\begin{center}\includegraphics[  width=7.4cm,
  angle=270,
  origin=lB]{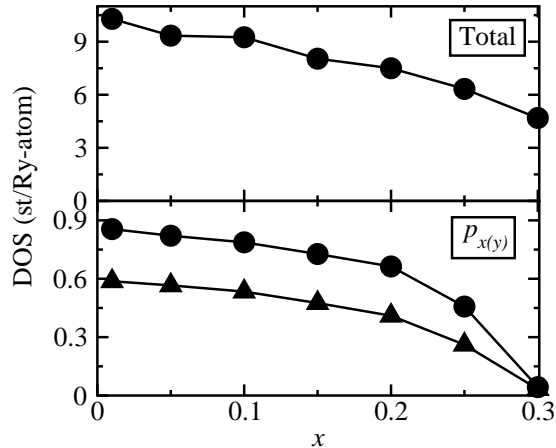}\end{center}

\caption{The calculated variation in the total (upper panel) and the $p_{x(y)}$
(lower panel) densities of states at the Fermi energy as a function
of $x$ in $Mg(B_{1-x}C_{x})_{2}$ alloys. In the lower panel, both
$B$ (filled circle) and $C$ (filled triangle) $p_{x(y)}$ densities
of states are shown.}
\end{figure}

\subsection{The superconducting transition temperature}

We have used the Gaspari-Gyorffy formalism to calculate the Hopfield
parameter and then the electron-phonon coupling constant $\lambda $
of $Mg(B_{1-x}C_{x})_{2}$ alloys. Our calculations show that the
$C$ atoms in $Mg(B_{1-x}C_{x})_{2}$ alloys couple to the phonons
as strongly as the $B$ atoms. For example, at $x=0.2$ we find $\lambda $
(per atom) for $B$ and $C$ to be $0.31$ and $0.24$, respectively.
It is interesting to note that as the $C$ $p_{x(y)}$ bands fill
up somewhat later than the $B$ $p_{x(y)}$ bands, the calculated
$\lambda $ of $C$ becomes  greater than that of $B$ for $x\sim 0.3$. 

To see how the changes in the electronic properties of $Mg(B_{1-x}C_{x})_{2}$
alloys affect the superconducting transition temperature, we show
in Fig. 5 the calculated $T_{c}$ as a function of $x$ in $Mg(B_{1-x}C_{x})_{2}$
alloys. The experimentally observed $T_{c}$ values as reported in
Refs. \cite{bharathi,ribeiro} are also shown in Fig. 5. Note that
for $x=0.1$, Refs. \cite{bharathi} and \cite{ribeiro} find $T_{c}$
to be equal to $34\, K$ and $22\, K$, respectively. Our calculations
show that the $T_{c}$ drops to $\sim 0\, K$ from $20\, K$ as $x$
changes from $0.2$ to 0.3, which is consistent with the filling up
of the $p_{x(y)}$ bands as discussed above. The calculated trend
in the variation of $T_{c}$ in $Mg(B_{1-x}C_{x})_{2}$ alloys is
in qualitative agreement with the experimental results as shown in
Fig. 5. 

Some of the differences between the calculated (more accurately than
the present approach) and the observed $T_{c}$ in $Mg(B_{1-x}C_{x})_{2}$
alloys may be due to an inaccurate determination of the $C$ content
in these alloys, as pointed out in Ref. \cite{ribeiro}. In the present
case, the calculation of $T_{c}$ incorporates two additional approximations,
(i) use of a constant phonon frequency over the whole range of $x$
and (ii) use of Gaspari-Gyorffy formalism for estimating the electron-phonon
matrix elements, which is known to underestimate the strength of the
coupling. However, we expect the trend in the variation of $T_{c}$
in $Mg(B_{1-x}C_{x})_{2}$ alloys, as shown in Fig. 5, to be reliable.

\begin{figure}
\begin{center}\includegraphics[  width=7.4cm,
  angle=270,
  origin=lB]{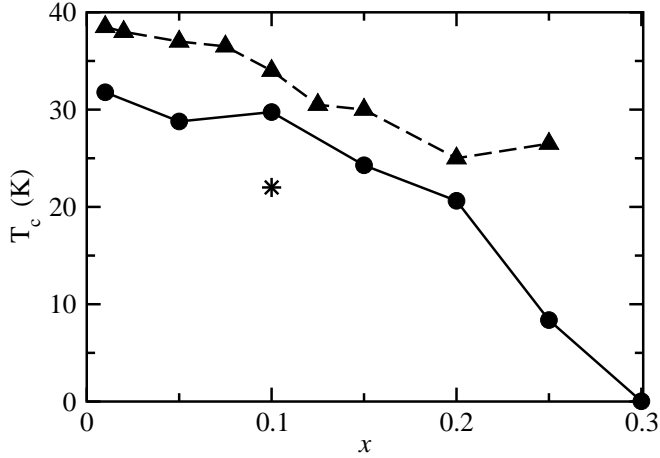}\end{center}

\caption{The calculated (filled circle) variation of $T_{c}$ as a function
of concentration $x$ in $Mg(B_{1-x}C_{x})_{2}$ alloys. The experimental
results of Ref. \cite{bharathi} (filled triangle) and Ref. \cite{ribeiro}
(star) are also shown.}
\end{figure}

\section{CONCLUSIONS}

We have studied the changes in the electronic structure of $Mg(B_{1-x}C_{x})_{2}$
alloys as a function of $x$ with $0\leq x\leq 0.3$ using the KKR-ASA
CPA method. We find that the changes in the electronic structure of
$Mg(B_{1-x}C_{x})_{2}$ alloys , especially near the Fermi energy
$E_{F}$, come mainly from the outward movement of $E_{F}$ with increasing
$x$ and the effects of disorder in the $B$ plane are small. In particular,
our results show a sharp decline in $B$ and $C$ $p_{x(y)}$ states
for $0.2\leq $$x$$\leq 0.3$. Our calculated variation in $T_{c}$
of $Mg(B_{1-x}C_{x})_{2}$ alloys is in qualitative agreement with
the experiments.


\begin{thebibliography}{10}
\expandafter\ifx\csname url\endcsname\relax
  \def\url#1{\texttt{#1}}\fi
\expandafter\ifx\csname urlprefix\endcsname\relax\def\urlprefix{URL }\fi

\bibitem{felner}
I.~Felner, Physica C 353 (2001) 11.

\bibitem{zhao}
Y.~G. Zhao, X.~P. Zhang, P.~T. Qiao, H.~T. Zhang, S.~L. Jia, B.~S. Cao, M.~H.
  Zhu, Z.~H. Han, X.~L. Wang, B.~L. Gu, Physica C 361 (2001) 91.

\bibitem{xiang}
J.~Y. Xiang, D.~N. Zheng, J.~Q. Li, L.~Li, P.~L. Lang, H.~Chen, C.~Dong, G.~C.
  Che, Z.~A. Ren, H.~H. Qi, H.~Y. Tian, Y.~M. Ni, Z.~X. Zhao, cond-mat/0104366.

\bibitem{slusky}
J.~S. Slusky, N.~Rogado, K.~A. Regan, M.~A. Hayward, P.~Khalifah, T.~He,
  K.~Inumaru, S.~M. Loureiro, M.~K. Haas, H.~W. Zandbergen, R.~J. Cava, Nature
  410 (2001) 343.

\bibitem{cimberle}
M.~R. Cimberle, M.~Novak, P.~Manfrinetti, A.~Palenzona, cond-mat/0105212.

\bibitem{moritomo}
Y.~Moritomo, S.~Xu, cond-mat/0104568.

\bibitem{buzea}
C.~Buzea, T.~Yamashita, Superconductors, Science and Technology 14 (2001) R115.

\bibitem{ahn}
J.~S. Ahn, E.~J. Choi, cond-mat/0103069.

\bibitem{paran}
M.~Paranthaman, J.~F. Thompson, D.~K. Christen, Physica C 355 (2001) 1.

\bibitem{takenobu}
T.~Takenobu, T.~Ito, D.~H. Chi, K.~Prassides, Y.~Iwasa, Phys.\ Rev. B 64 (2001)
  134513.

\bibitem{zhang}
S.~Zhang, J.~Zhang, T.~Zhao, C.~Rong, B.~Shen, Z.~Cheng, cond-mat/0103203.

\bibitem{bharathi}
A.~Bharathi, S.~J. Balaselvi, S.~Kalavathi, G.~L.~N. Reddy, V.~S. Sastry,
  Y.~Haritharan, T.~S. Radhakrishnan, Physica C 370 (2002) 211.

\bibitem{ribeiro}
R.~A. Ribeiro, S.~Bud'ko, C.~Petrovic, P.~C. Canfield, Physica C 382 (2002)
  166.

\bibitem{mehl}
M.~J. Mehl, D.~A. Papaconstantopoulos, D.~J. Singh, Phys.\ Rev. B 64 (2001)
  134513.

\bibitem{pps_cpa}
P.~P. Singh, A.~Gonis, Phys.\ Rev. B 49 (1994) 1642.

\bibitem{faulkner}
J.~Faulkner, Prog. Mat. Sci. 27 (1982) 1.

\bibitem{gaspari}
G.~Gaspari, B.~L. Gyorffy, Phys.\ Rev. Lett. 28 (1972) 801.

\bibitem{allen1}
P.~B. Allen, R.~C. Dynes, Phys.\ Rev. B 12 (1975) 905.

\bibitem{pps_mgtab2}
P.~J.~T. Joseph, P.~P. Singh, Solid State Commun. 121 (2002) 467.

\bibitem{pps_mgalb2}
P.~P. Singh, Physica C 382 (2002) 381.

\bibitem{pps_3dmgb2}
P.~P. Singh, P.~J.~T. Joseph, J. Phys.:Condens. Matter 14 (2001) 12441.

\bibitem{perdew1}
J.~P. Perdew, Y.~Wang, Phys.\ Rev. B 45 (1992) 13244.

\bibitem{perdew2}
J.~P. Perdewa, K.~Burke, M.~Ernzerhof, Phys.\ Rev. Lett. 77 (1996) 3865.

\bibitem{kong}
Y.~Kong, O.~V. Dolgov, O.~Jepsen, O.~K. Andersen, Phys.\ Rev. B 64 (2001)
  20501.

\bibitem{bohnen}
K.-P. Bohnen, R.~Heid, B.~Renker, Phys.\ Rev. Lett. 86 (2001) 5771.

\bibitem{choi}
H.~J. Choi, D.~Roundy, H.~Sun, M.~L. Cohen, S.~G. Louie, Phys.\ Rev. B 66
  (2002) 020513.

\bibitem{pps_nbb2}
P.~P. Singh, Solid State Commun. 125 (2003) 323.

\bibitem{pps_mgb2_prl}
P.~P. Singh, Phys.\ Rev. Lett. 87 (2001) 087004.

\bibitem{kortus}
J.~Kortus, I.~I. Mazin, K.~D. Belashchenko, V.~P. Antropov, L.~L. Boyer, Phys.\
  Rev. Lett. 86 (2001) 4656.

\bibitem{zhu}
Y.~Zhu, A.~R. Moodenbaugh, G.~Schneider, J.~W. Davenport, T.~Vogt, Q.~Li,
  G.~Gu, D.~A. Fischer, J.~Tafto, Phys.\ Rev. Lett. 88 (2002) 247002.

\bibitem{klie}
R.~F. Klie, Y.~Zhu, G.~Schneider, J.~Tafto, cond-mat/0301575.

\end{thebibliography}
\end{document}